# The Pythagorean Won-Loss Formula and Hockey:
## A Statistical Justification for Using the Classic Baseball Formula as an Evaluative Tool in Hockey

By Kevin D. Dayaratna and Steven J. Miller[*]

October 16, 2013

## Abstract

Originally devised for baseball, the Pythagorean Won-Loss formula estimates the percentage of games a team should have won at a particular point in a season. For decades, this formula had no mathematical justification. In 2006, Steven Miller provided a statistical derivation by making some heuristic assumptions about the distributions of runs scored and allowed by baseball teams. We make a similar set of assumptions about hockey teams and show that the formula is just as applicable to hockey as it is to baseball. We hope that this work spurs research in the use of the Pythagorean Won-Loss formula as an evaluative tool for sports outside baseball.

## I. Introduction

The Pythagorean Won-Loss formula has been around for decades. Initially devised by the well-known baseball statistician Bill James during the early 1980s, the Pythagorean Won-Loss formula provides the winning percentage (WP) a baseball team should be expected to have at a particular time during a season based on its runs scored (RS) and allowed (RA):

$$WP \approx \frac{RS^{\gamma}}{RS^{\gamma} + RA^{\gamma}}.$$

Early on, James believed the exponent to be two (thus the name "Pythagorean" from a sum of squares). Empirical examination later advised that $\gamma \approx 1.8$ was more suitable.

For years, baseball statisticians used the Pythagorean Won-Loss formula to predict a team's won-loss record at the end of the season. "Sabermetricians" (statistical analysts affiliated with the Society of American Baseball Research) also used the percentage to comment on a team's level of over-performance/under-performance as well as the value of adding certain players to their lineup. Until recently, however, the Pythagorean Won-Loss formula had been devoid of any theoretical justification from first principles. Miller (2007) addressed this issue by assuming that RS and RA follow independent Weibull probability distributions and subsequently derived James's formula by computing the probability that the runs a team scores exceeds the runs it allows. He found, as empirical observation had consistently suggested, that the most suitable value of $\gamma$ was indeed approximately 1.8.

---

[*] Kevin D. Dayaratna (kevind@math.umd.edu) is a doctoral candidate in Mathematical Statistics at the University of Maryland. Steven J. Miller (steven.j.miller@williams.edu) is an Associate Professor of Mathematics and Statistics at Williams College. The second author is supported by NSF grant DMS1265673 and NSF grant DMS0970067. We would like to thank Eric Fritz for writing a Java script to download hockey data from ESPN.com, Mike Taylor, Hovannes Abramyan and Arnal Dayaratna for proofreading; Armaan Ambati for help with the data; and Benjamin Kedem for insightful comments. An abridged version of this manuscript has been accepted for publication by *The Hockey Research Journal* of the Society for International Hockey Research.

A few researchers have applied Bill James's model to other sports. For example, Schatz (2003) applied the model to football and determined that an appropriate value of $\gamma$ is around 2.37. Oliver (2004) did the same for basketball and determined that an appropriate value of $\gamma$ is around 14. Rosenfeld et al. (2010) drew upon this research and used the Pythagorean Won-Loss formula to predict overtime wins in baseball, basketball, and football.

Cochran and Blackstock (2009) applied the Pythagorean Won-Loss formula to hockey, as have Chris Apple and Marc Foster (Apple and Foster 2002; Foster 2010). Cochran and Blackstock used least squares estimation to estimate James's model as well as several modifications of it. They found that James's original Pythagorean Won-Loss formula, with a value of $\gamma$ around 1.927, is just as accurate as the results produced by more complex models.

Few outside of Alan Ryder (hockeyanalytics.com), however, have provided a theoretical verification from first principles for applying the Pythagorean Won-Loss formula to any sport other than baseball. We add to his efforts here. Specifically, we make the same assumptions that Miller (2007) made for baseball and find that the Pythagorean Won-Loss formula applies just as well to hockey as it does to baseball. Our results thus provide theoretical justification for using the Pythagorean Won-Loss formula, initially intended for baseball, as an evaluative tool in hockey.

Our work is organized as follows. We first discuss our model and estimation results; in particular, we sketch the derivation of the Pythagorean Won-Loss formula. Afterwards, we examine our model's statistical validity by performing tests of statistical independence as well as goodness of fit. Finally, we conclude by summarizing our findings and discussing potential avenues of future research.

## II. Model Development

In this section, we prove that if GS and GA are drawn from independent translated Weibull distributions then the Pythagorean Won-Loss formula holds. Specifically, we assume that the distribution of the number of goals a hockey team scores and the number of goals it allows each follow independent translated two-parameter Weibull distributions with the following probability density functions:

$$f(x;\alpha_{GS},\gamma) = \frac{\gamma}{\alpha_{GS}}(\frac{x+0.5}{\alpha_{GS}})^{\gamma-1}e^{-\left(\frac{x+0.5}{\alpha_{GS}}\right)^{\gamma}}I(x>-0.5)$$

$$f(y;\alpha_{GA},\gamma) = \frac{\gamma}{\alpha_{GA}}(\frac{y+0.5}{\alpha_{GA}})^{\gamma-1}e^{-\left(\frac{y+0.5}{\alpha_{GA}}\right)^{\gamma}}I(y>-0.5)$$

where $I(x>-0.5)$ and $I(y>-0.5)$ are indicator variables that are equal to 1 if their arguments are greater than -0.5 and are zero otherwise. We specifically translated the Weibull densities by a factor of 0.5 to ensure that our data (the integer representing the score) is at the center of the bins for our chi-squared goodness of fit tests. Continuous distributions are used to facilitate computation by transforming sums into integrals, and facilitate getting a simple, closed-form expression such as the Pythagorean formula. Of course, continuous distributions do not truly represent reality as baseball and hockey teams only score integral values of points; however, the Weibull is a flexible distribution and by appropriately choosing its parameters, it can fit many data sets. Miller (2007) showed the Pythagorean Won-Loss formula can be derived by computing the probability that the number of goals a team scores is greater than the number of goals it allows. We sketch the argument below:

$$\text{Pythag\_WL} = \Pr(x > y)$$

$$= \int_{-0.5}^{\infty} \int_{-0.5}^{x} f(x; \alpha_{GS}, \gamma) \cdot f(y; \alpha_{GA}, \gamma) dy dx$$

$$= \int_{-0.5}^{\infty} \int_{-0.5}^{x} \frac{\gamma}{\alpha_{GS}} (\frac{x+0.5}{\alpha_{GS}})^{\gamma-1} e^{-\left(\frac{x+0.5}{\alpha_{GS}}\right)^{\gamma}} \frac{\gamma}{\alpha_{GA}} (\frac{y+0.5}{\alpha_{GA}})^{\gamma-1} e^{-\left(\frac{y+0.5}{\alpha_{GA}}\right)^{\gamma}} dy dx$$

$$= \int_{0}^{\infty} \frac{\gamma}{\alpha_{GS}} (\frac{x}{\alpha_{GS}})^{\gamma-1} e^{-\left(\frac{x}{\alpha_{GS}}\right)^{\gamma}} \left[ \int_{0}^{x} \frac{\gamma}{\alpha_{GA}} (\frac{y}{\alpha_{GA}})^{\gamma-1} e^{-\left(\frac{y}{\alpha_{GA}}\right)^{\gamma}} dy \right] dx$$

$$= \int_{0}^{\infty} \frac{\gamma}{\alpha_{GS}} (\frac{x}{\alpha_{GS}})^{\gamma-1} e^{-\left(\frac{x}{\alpha_{GS}}\right)^{\gamma}} \left[ 1 - e^{-\left(\frac{x}{\alpha_{GA}}\right)^{\gamma}} \right] dx$$

$$\text{Pythag\_WL} = \frac{\alpha_{GS}^{\gamma}}{\alpha_{GS}^{\gamma} + \alpha_{GA}^{\gamma}}.$$

The mean goals scored (*GS*) and mean goals allowed (*GA*) for our translated Weibull densities are: $GS = \alpha_{GS}\Gamma(1 + \gamma^{-1}) - 0.5$ and $GA = \alpha_{GA}\Gamma(1 + \gamma^{-1}) - 0.5$ (Casella and Berger 2002; Miller 2006). Therefore, after a bit of algebra:

$$\text{Pythag\_WL} = \frac{(GS + 0.5)^{\gamma}}{(GS + 0.5)^{\gamma} + (GA + 0.5)^{\gamma}}.$$

Maximum likelihood parameter estimation of our Weibull densities enables us to compute these Pythagorean expectations.

### III. Data and Results

We compiled data (goals scored and goals allowed) from ESPN.com for each of the 30 NHL teams over the course of the 2008/09, 2009/10, and 2010/11 regular seasons. We estimated our parameters simultaneously via maximum likelihood estimation (MLE). We also performed tests of statistical independence as well as goodness of fit tests. Figures 1 through 4 are some representative plots of the observed data and the best fit Weibulls for the 2010/11 season. The complete plots are available from the authors. We have chosen the 2011 Stanley Cup champions, the Boston Bruins, their opponent, the Vancouver Canucks, the New Jersey Devils (whose 38 wins, 39 losses and 5 overtime losses makes them close to an average team), and the Edmonton Oilers, who had the worst record in 2010/11:

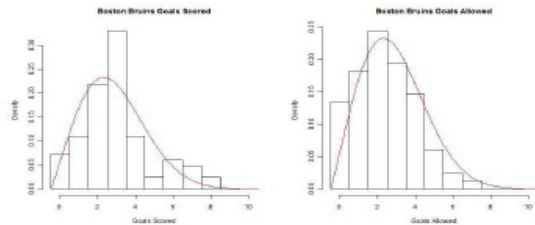

FIGURE 1. Boston Bruins: Goals scored and allowed, 2011.

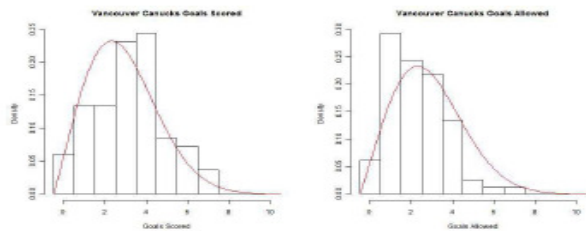

FIGURE 2. Vancouver Canucks: Goals scored and allowed, 2011.

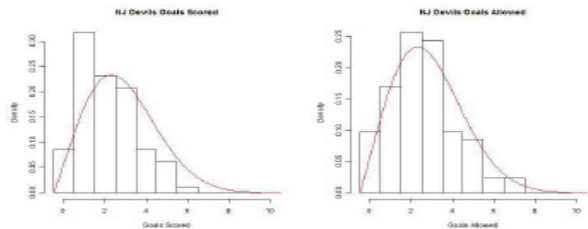

FIGURE 3. New Jersey Devils: Goals scored and allowed, 2011.

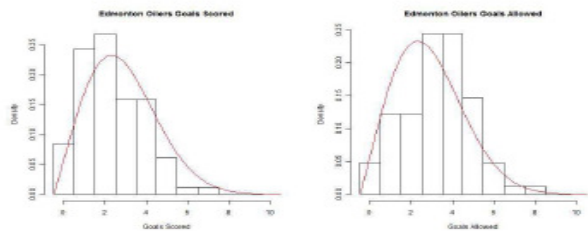

FIGURE 4. Edmonton Oilers: Goals scored and allowed, 2011.

Our results from our maximum likelihood estimation, our computation of each of the 30 NHL team's Pythagorean won loss formula (Pythag_WL), and our computed difference between the observed number of games won and the expected number of games won (Diff), are below:

2008/09 National Hockey League Eastern Conference

| Team | Games Won | Games Lost | Actual WL | Pythag_WL | Diff | $\gamma$ | $\alpha_{GS}$ | $\alpha_{GA}$ |
|---|---|---|---|---|---|---|---|---|
| Boston Bruins | 53 | 29 | 0.646 | 0.639 | 0.57 | 2.11 | 4.31 | 3.28 |
| NJ Devils | 51 | 31 | 0.622 | 0.565 | 4.71 | 1.99 | 3.91 | 3.43 |
| Washington Capitals | 50 | 32 | 0.610 | 0.534 | 6.25 | 2.31 | 4.24 | 4.00 |
| Carolina Hurricanes | 45 | 37 | 0.549 | 0.534 | 1.22 | 2.12 | 3.89 | 3.65 |
| Pittsburgh Penguins | 45 | 37 | 0.549 | 0.551 | -0.16 | 2.24 | 4.21 | 3.84 |
| Philadelphia Flyers | 44 | 38 | 0.537 | 0.567 | -2.46 | 2.37 | 4.25 | 3.79 |
| New York Rangers | 43 | 39 | 0.524 | 0.466 | 4.79 | 2.02 | 3.39 | 3.63 |
| Buffalo Sabres | 41 | 41 | 0.500 | 0.531 | -2.55 | 2.17 | 4.00 | 3.78 |
| Florida Panthers | 41 | 41 | 0.500 | 0.506 | -0.46 | 2.12 | 3.78 | 3.74 |
| Montreal Canadiens | 41 | 41 | 0.500 | 0.511 | -0.86 | 2.45 | 4.01 | 3.94 |
| Ottawa Senators | 36 | 46 | 0.439 | 0.454 | -1.27 | 2.27 | 3.54 | 3.84 |
| Atlanta Thrashers | 35 | 47 | 0.427 | 0.469 | -3.46 | 2.31 | 4.13 | 4.36 |
| Toronto Maple Leafs | 34 | 48 | 0.415 | 0.442 | -2.24 | 2.27 | 4.08 | 4.53 |
| New York Islanders | 26 | 56 | 0.317 | 0.339 | -1.81 | 2.25 | 3.30 | 4.44 |
| Tampa Bay Lightning | 24 | 58 | 0.293 | 0.378 | -6.96 | 2.31 | 3.50 | 4.34 |

2008/09 National Hockey League Western Conference

| Team | Games Won | Games Lost | Actual WL | Pythag_WL | Diff | $\gamma$ | $\alpha_{GS}$ | $\alpha_{GA}$ |
|---|---|---|---|---|---|---|---|---|
| San Jose Sharks | 53 | 29 | 0.646 | 0.580 | 5.45 | 2.07 | 4.02 | 3.44 |
| Detroit Red Wings | 51 | 31 | 0.622 | 0.558 | 5.22 | 2.29 | 4.46 | 4.03 |
| Calgary | 46 | 36 | 0.561 | 0.508 | 4.36 | 2.11 | 4.05 | 3.99 |

| Team | Games Won | Games Lost | Actual WL | Pythag_WL | Diff | $\gamma$ | $\alpha_{GS}$ | $\alpha_{GA}$ |
|---|---|---|---|---|---|---|---|---|
| Flames | | | | | | | | |
| Chicago Blackhawks | 46 | 36 | 0.561 | 0.572 | -0.87 | 2.09 | 4.12 | 3.59 |
| Vancouver Canucks | 45 | 37 | 0.549 | 0.536 | 1.03 | 2.08 | 3.89 | 3.63 |
| Anaheim Ducks | 42 | 40 | 0.512 | 0.510 | 0.17 | 2.25 | 3.91 | 3.84 |
| Columbus Blue Jackets | 41 | 41 | 0.500 | 0.484 | 1.31 | 1.99 | 3.63 | 3.75 |
| St Louis Blues | 41 | 41 | 0.500 | 0.492 | 0.62 | 2.16 | 3.74 | 3.79 |
| Minnesota Wild | 40 | 42 | 0.488 | 0.555 | -5.50 | 2.12 | 3.62 | 3.27 |
| Nashville Predators | 40 | 42 | 0.488 | 0.462 | 2.12 | 1.94 | 3.48 | 3.77 |
| Edmonton Oilers | 38 | 44 | 0.463 | 0.474 | -0.83 | 2.09 | 3.79 | 3.98 |
| Dallas Stars | 36 | 46 | 0.439 | 0.474 | -2.83 | 2.09 | 3.82 | 4.02 |
| Phoenix Coyotes | 36 | 46 | 0.439 | 0.423 | 1.31 | 2.00 | 3.44 | 4.01 |
| LA Kings | 34 | 48 | 0.415 | 0.469 | -4.45 | 1.97 | 3.47 | 3.70 |
| Colorado Avalanche | 32 | 50 | 0.390 | 0.418 | -2.26 | 2.00 | 3.39 | 4.00 |

2009/10 National Hockey League Eastern Conference

| Team | Games Won | Games Lost | Actual WL | Pythag_WL | Diff | $\gamma$ | $\alpha_{GS}$ | $\alpha_{GA}$ |
|---|---|---|---|---|---|---|---|---|
| Washington Capitals | 54 | 28 | 0.659 | 0.635 | 1.93 | 2.57 | 4.80 | 3.87 |
| NJ Devils | 48 | 34 | 0.585 | 0.56 | 2.08 | 2.10 | 3.60 | 3.21 |
| Buffalo Sabres | 45 | 37 | 0.549 | 0.571 | -1.81 | 2.21 | 3.84 | 3.37 |
| Pittsburgh Penguins | 47 | 35 | 0.573 | 0.548 | 2.08 | 2.18 | 4.14 | 3.79 |
| Ottawa Senators | 44 | 38 | 0.537 | 0.471 | 5.40 | 2.14 | 3.65 | 3.85 |
| Boston Bruins | 39 | 43 | 0.476 | 0.515 | -3.28 | 1.99 | 3.41 | 3.30 |
| Philadelphia Flyers | 41 | 41 | 0.5 | 0.522 | -1.82 | 1.94 | 3.82 | 3.65 |
| Montreal | 39 | 43 | 0.476 | 0.489 | -1.13 | 2.18 | 3.55 | 3.62 |

| Team | Games Won | Games Lost | Actual WL | Pythag_WL | Diff | $\gamma$ | $\alpha_{GS}$ | $\alpha_{GA}$ |
|---|---|---|---|---|---|---|---|---|
| Canadiens | | | | | | | | |
| New York Rangers | 38 | 44 | 0.463 | 0.512 | -3.96 | 1.95 | 3.64 | 3.55 |
| Atlanta Thrashers | 35 | 47 | 0.427 | 0.468 | -3.40 | 2.25 | 3.82 | 4.04 |
| Carolina Hurricanes | 35 | 47 | 0.427 | 0.471 | -3.65 | 2.29 | 3.81 | 4.00 |
| Tampa Bay Lightning | 34 | 48 | 0.415 | 0.414 | 0.04 | 2.13 | 3.54 | 4.16 |
| New York Islanders | 34 | 48 | 0.415 | 0.424 | -0.75 | 2.21 | 3.63 | 4.18 |
| Florida Panthers | 32 | 50 | 0.39 | 0.449 | -4.81 | 1.97 | 3.47 | 3.85 |
| Toronto Maple Leafs | 30 | 52 | 0.366 | 0.407 | -3.41 | 2.30 | 3.55 | 4.18 |

## 2009/10 National Hockey League Western Conference

| Team | Games Won | Games Lost | Actual WL | Pythag_WL | Diff | $\gamma$ | $\alpha_{GS}$ | $\alpha_{GA}$ |
|---|---|---|---|---|---|---|---|---|
| San Jose Sharks | 51 | 31 | 0.622 | 0.579 | 3.51 | 2.23 | 4.14 | 3.59 |
| Chicago Blackhawks | 52 | 30 | 0.634 | 0.587 | 3.86 | 2.15 | 4.16 | 3.53 |
| Vancouver Canucks | 49 | 33 | 0.598 | 0.573 | 1.97 | 2.22 | 4.22 | 3.69 |
| Phoenix Coyotes | 50 | 32 | 0.610 | 0.545 | 5.33 | 2.17 | 3.64 | 3.35 |
| Detroit Red Wings | 44 | 38 | 0.537 | 0.532 | 0.37 | 2.15 | 3.73 | 3.51 |
| LA Kings | 46 | 36 | 0.561 | 0.560 | 0.12 | 2.24 | 3.93 | 3.54 |
| Nashville Predators | 47 | 35 | 0.573 | 0.501 | 5.95 | 2.14 | 3.65 | 3.65 |
| Colorado Avalanche | 43 | 39 | 0.524 | 0.498 | 2.19 | 2.25 | 3.82 | 3.84 |
| St Louis Blues | 40 | 42 | 0.488 | 0.498 | -0.84 | 2.18 | 3.64 | 3.65 |
| Calgary Flames | 40 | 42 | 0.488 | 0.484 | 0.30 | 2.01 | 3.36 | 3.47 |
| Anaheim Ducks | 39 | 43 | 0.476 | 0.484 | -0.66 | 2.35 | 3.86 | 3.97 |

| Team | Games Won | Games Lost | Actual WL | Pythag_WL | Diff | $\gamma$ | $\alpha_{GS}$ | $\alpha_{GA}$ |
|---|---|---|---|---|---|---|---|---|
| Dallas Stars | 37 | 45 | 0.451 | 0.476 | -2.03 | 2.42 | 3.85 | 4.01 |
| Minnesota Wild | 38 | 44 | 0.463 | 0.450 | 1.12 | 2.50 | 3.60 | 3.91 |
| Columbus Blue Jackets | 32 | 50 | 0.390 | 0.408 | -1.48 | 2.12 | 3.50 | 4.17 |
| Edmonton Oilers | 27 | 55 | 0.329 | 0.377 | -3.87 | 2.35 | 3.55 | 4.40 |

## 2010/11 National Hockey League Eastern Conference

| Team | Games Won | Games Lost | Actual WL | Pythag_WL | Diff | $\gamma$ | $\alpha_{GS}$ | $\alpha_{GA}$ |
|---|---|---|---|---|---|---|---|---|
| Pittsburgh Penguins | 49 | 33 | 0.598 | 0.569 | 2.34 | 2.00 | 3.82 | 3.32 |
| Washington Capitals | 48 | 34 | 0.585 | 0.560 | 2.09 | 1.91 | 3.67 | 3.23 |
| Philadelphia Flyers | 47 | 35 | 0.573 | 0.572 | 0.12 | 2.14 | 4.15 | 3.62 |
| Boston Bruins | 46 | 36 | 0.561 | 0.586 | -2.05 | 1.89 | 3.91 | 3.26 |
| Tampa Bay Lightning | 46 | 36 | 0.561 | 0.493 | 5.55 | 2.00 | 3.89 | 3.94 |
| Montreal Canadiens | 44 | 38 | 0.537 | 0.504 | 2.64 | 1.93 | 3.49 | 3.46 |
| New York Rangers | 44 | 38 | 0.537 | 0.571 | -2.83 | 1.88 | 3.79 | 3.25 |
| Buffalo Sabres | 43 | 39 | 0.524 | 0.531 | -0.57 | 2.14 | 3.93 | 3.71 |
| Carolina Hurricanes | 40 | 42 | 0.488 | 0.503 | -1.26 | 2.17 | 3.84 | 3.82 |
| NJ Devils | 38 | 44 | 0.463 | 0.426 | 3.09 | 1.96 | 2.95 | 3.44 |
| Toronto Maple Leafs | 37 | 45 | 0.451 | 0.464 | -1.04 | 2.09 | 3.65 | 3.91 |
| Atlanta Thrashers | 34 | 48 | 0.415 | 0.404 | 0.90 | 2.32 | 3.62 | 4.28 |
| Ottawa Senators | 32 | 50 | 0.390 | 0.386 | 0.36 | 2.07 | 3.20 | 4.01 |
| Florida Panthers | 30 | 52 | 0.366 | 0.442 | -6.21 | 2.31 | 3.29 | 3.64 |
| New York Islanders | 30 | 52 | 0.366 | 0.455 | -7.32 | 2.14 | 3.79 | 4.12 |

2010/11 National Hockey League Western Conference

| Team | Games Won | Games Lost | Actual WL | Pythag_WL | Diff | $\gamma$ | $\alpha_{GS}$ | $\alpha_{GA}$ |
|------|-----------|------------|-----------|-----------|------|----------|---------------|---------------|
| Vancouver Canucks | 54 | 28 | 0.659 | 0.644 | 1.20 | 2.15 | 4.13 | 3.14 |
| San Jose Sharks | 48 | 34 | 0.585 | 0.562 | 1.88 | 2.21 | 3.94 | 3.51 |
| Detroit Red Wings | 47 | 35 | 0.573 | 0.541 | 2.61 | 2.24 | 4.16 | 3.86 |
| Anaheim Ducks | 47 | 35 | 0.573 | 0.500 | 5.96 | 2.11 | 3.82 | 3.82 |
| LA Kings | 46 | 36 | 0.561 | 0.526 | 2.91 | 1.98 | 3.52 | 3.34 |
| Chicago Blackhawks | 44 | 38 | 0.537 | 0.558 | -1.77 | 2.29 | 4.08 | 3.68 |
| Nashville Predators | 44 | 38 | 0.537 | 0.549 | -0.98 | 2.15 | 3.55 | 3.24 |
| Phoenix Coyotes | 43 | 39 | 0.524 | 0.495 | 2.44 | 2.16 | 3.68 | 3.71 |
| Dallas Stars | 42 | 40 | 0.512 | 0.464 | 3.94 | 2.23 | 3.61 | 3.85 |
| Calgary Flames | 41 | 41 | 0.500 | 0.524 | -1.96 | 2.10 | 4.00 | 3.82 |
| Minnesota Wild | 39 | 43 | 0.476 | 0.450 | 2.13 | 2.03 | 3.40 | 3.76 |
| St Louis Blues | 38 | 44 | 0.463 | 0.497 | -2.78 | 1.94 | 3.81 | 3.83 |
| Columbus Blue Jackets | 34 | 48 | 0.415 | 0.408 | 0.50 | 2.25 | 3.49 | 4.12 |
| Colorado Avalanche | 30 | 52 | 0.366 | 0.423 | -4.70 | 2.42 | 3.83 | 4.35 |
| Edmonton Oilers | 25 | 57 | 0.305 | 0.374 | -5.64 | 2.16 | 3.29 | 4.17 |

The maximum likelihood estimated value of $\gamma$ is almost always slightly above 2, averaging 2.15 for the 2008/09 season (standard deviation 0.133), 2.19 for the 2009/10 season (standard deviation 0.14), and 2.10 (standard deviation 0.144) for the 2010/11 season, which is reasonably close to the estimates computed in Cochran and Blackstock (2009). Our results also indicate that many of the top teams, including the Washington Capitals, NJ Devils, San Jose Sharks, and the Chicago Blackhawks and Vancouver Canucks performed better than expected over the course of the seasons examined.

In the next two sections, we test the fundamental assumptions our model makes – namely statistical independence between goals scored and goals allowed and the appropriateness of the Weibull densities to model our data.

IV. Model Testing: Statistical Independence of Goals Scored and Goals Allowed

Naively, one would think that the distributions of goals scored and goals allowed should be treated as dependent distributions. For example, if a team has a big lead, the coaching staff might change players or use up remaining time on the clock. On the other hand, if a team is trailing toward the end of a game, the staff may pull the goalie to increase the probability of scoring.

Some of these arguments also apply to other sports, including baseball. Recent research in "sabermetrics" (Ciccolella 2006; Miller 2007), however, suggests that the distributions of runs scored and runs allowed can indeed be considered independent. We tested whether this argument is true for hockey by performing non-parametric statistical tests of Kendall's Tau and Spearman's Rho (Hogg et al 2005) for each team on a game-by-game basis. Below are our results of each of these tests, which test the null hypothesis that the distributions of GS and GA are independent:

Tests of Kendall's Tau and Spearman's Rho

| Team | Kendall's Tau for 2008/09 Season | p-value for 2008/09 Season | Kendall's Tau for 2009/10 Season | p-value for 2009/10 Season | Kendall's Tau for 2010/11 Season | p-value for 2010/11 Season |
|---|---|---|---|---|---|---|
| Anaheim Ducks | 0.075 | 0.156 | -0.105 | 0.078 | 0.008 | 0.450 |
| Atlanta Thrashers | -0.023 | 0.381 | 0.027 | 0.356 | -0.061 | 0.205 |
| Boston Bruins | 0.126 | 0.044 | -0.047 | 0.264 | -0.108 | 0.072 |
| Buffalo Sabres | -0.123 | 0.049 | -0.063 | 0.197 | 0.083 | 0.131 |
| Calgary Flames | -0.056 | 0.227 | -0.055 | 0.228 | -0.031 | 0.340 |
| Carolina Hurricanes | -0.129 | 0.042 | -0.165 | 0.013 | -0.112 | 0.066 |
| Chicago Blackhawks | -0.048 | 0.261 | 0.060 | 0.211 | -0.056 | 0.227 |
| Colorado Avalanche | 0.036 | 0.313 | -0.042 | 0.287 | -0.047 | 0.264 |
| Columbus Blue Jackets | 0.042 | 0.285 | 0.063 | 0.200 | -0.090 | 0.113 |
| Dallas Stars | 0.049 | 0.253 | -0.089 | 0.116 | -0.126 | 0.045 |
| Detroit Red Wings | 0.006 | 0.466 | -0.009 | 0.453 | -0.003 | 0.484 |
| Edmonton Oilers | -0.042 | 0.288 | 0.017 | 0.412 | -0.217 | 0.002 |
| Florida Panthers | -0.105 | 0.078 | 0.003 | 0.485 | -0.082 | 0.135 |
| LA Kings | 0.073 | 0.164 | -0.017 | 0.409 | -0.008 | 0.455 |

| Minnesota Wild | -0.046 | 0.266 | -0.025 | 0.368 | -0.207 | 0.003 |
|---|---|---|---|---|---|---|
| Montreal Canadiens | -0.079 | 0.145 | -0.006 | 0.468 | -0.171 | 0.011 |
| Nashville Predators | 0.109 | 0.072 | 0.078 | 0.148 | -0.132 | 0.037 |
| New York Islanders | -0.019 | 0.399 | -0.056 | 0.224 | 0.017 | 0.409 |
| New York Rangers | 0.015 | 0.418 | -0.097 | 0.095 | -0.007 | 0.461 |
| NJ Devils | -0.089 | 0.114 | -0.096 | 0.099 | -0.125 | 0.046 |
| Ottawa Senators | 0.034 | 0.323 | -0.126 | 0.045 | -0.088 | 0.118 |
| Philadelphia Flyers | -0.038 | 0.303 | -0.023 | 0.376 | -0.097 | 0.095 |
| Phoenix Coyotes | -0.008 | 0.455 | -0.072 | 0.167 | -0.006 | 0.466 |
| Pittsburgh Penguins | -0.014 | 0.423 | -0.041 | 0.292 | -0.059 | 0.212 |
| San Jose Sharks | 0.083 | 0.131 | -0.125 | 0.047 | -0.047 | 0.262 |
| St Louis Blues | 0.032 | 0.332 | -0.032 | 0.332 | 0.030 | 0.346 |
| Tampa Bay Lightning | 0.100 | 0.089 | -0.065 | 0.190 | -0.026 | 0.364 |
| Toronto Maple Leafs | 0.031 | 0.337 | -0.043 | 0.280 | -0.037 | 0.308 |
| Vancouver Canucks | 0.047 | 0.264 | -0.130 | 0.040 | -0.088 | 0.118 |
| Washington Capitals | 0.025 | 0.368 | 0.023 | 0.381 | -0.036 | 0.316 |

| Team | Spearman's Rho for 2008-2009 Season | p-value for 2008-2009 Season | Spearman's Rho for 2009-2010 Season | p-value for 2009-2010 Season | Spearman's Rho for 2010-2011 Season | p-value for 2010-2011 Season |
|---|---|---|---|---|---|---|
| Anaheim Ducks | 0.145 | 0.193 | -0.123 | 0.272 | 0.030 | 0.789 |
| Atlanta Thrashers | -0.007 | 0.950 | 0.084 | 0.452 | -0.052 | 0.644 |
| Boston Bruins | 0.214 | 0.054 | -0.032 | 0.772 | -0.124 | 0.266 |
| Buffalo | -0.164 | 0.141 | -0.051 | 0.651 | 0.163 | 0.144 |

| | | | | | |
|---|---|---|---|---|---|
| Sabres | | | | | |
| Calgary Flames | -0.057 | 0.614 | -0.034 | 0.763 | -0.015 | 0.896 |
| Carolina Hurricanes | -0.152 | 0.172 | -0.217 | 0.050 | -0.116 | 0.300 |
| Chicago Blackhawks | -0.041 | 0.717 | 0.117 | 0.296 | -0.048 | 0.671 |
| Colorado Avalanche | 0.087 | 0.436 | -0.026 | 0.819 | -0.018 | 0.875 |
| Columbus Blue Jackets | 0.090 | 0.420 | 0.131 | 0.240 | -0.086 | 0.442 |
| Dallas Stars | 0.101 | 0.367 | -0.102 | 0.363 | -0.153 | 0.169 |
| Detroit Red Wings | 0.047 | 0.673 | 0.028 | 0.805 | 0.025 | 0.820 |
| Edmonton Oilers | -0.024 | 0.833 | 0.058 | 0.607 | -0.290 | 0.008 |
| Florida Panthers | -0.126 | 0.259 | 0.036 | 0.748 | -0.071 | 0.526 |
| LA Kings | 0.139 | 0.212 | 0.016 | 0.889 | 0.021 | 0.853 |
| Minnesota Wild | -0.037 | 0.738 | 0.001 | 0.992 | -0.276 | 0.012 |
| Montreal Canadiens | -0.085 | 0.447 | 0.030 | 0.791 | -0.223 | 0.044 |
| Nashville Predators | 0.186 | 0.094 | 0.140 | 0.210 | -0.166 | 0.136 |
| New York Islanders | 0.006 | 0.959 | -0.049 | 0.663 | 0.051 | 0.652 |
| New York Rangers | 0.062 | 0.579 | -0.120 | 0.283 | 0.019 | 0.867 |
| NJ Devils | -0.104 | 0.353 | -0.117 | 0.297 | -0.166 | 0.136 |
| Ottawa Senators | 0.079 | 0.481 | -0.177 | 0.111 | -0.100 | 0.372 |
| Philadelphia Flyers | -0.026 | 0.820 | -0.012 | 0.913 | -0.090 | 0.424 |
| Phoenix Coyotes | 0.012 | 0.918 | -0.064 | 0.569 | 0.033 | 0.765 |
| Pittsburgh Penguins | 0.001 | 0.996 | -0.030 | 0.791 | -0.048 | 0.666 |
| San Jose Sharks | 0.153 | 0.169 | -0.146 | 0.189 | -0.032 | 0.774 |
| St Louis Blues | 0.083 | 0.458 | -0.015 | 0.895 | 0.074 | 0.511 |
| Tampa Bay Lightning | 0.182 | 0.102 | -0.063 | 0.573 | 0.001 | 0.996 |
| Toronto Maple Leafs | 0.075 | 0.505 | -0.027 | 0.812 | -0.034 | 0.765 |
| Vancouver Canucks | 0.093 | 0.404 | -0.175 | 0.116 | -0.092 | 0.413 |
| Washington Capitals | 0.062 | 0.583 | 0.071 | 0.528 | -0.010 | 0.928 |

After we assume commonly-accepted critical thresholds of 0.05 and 0.10, instituting Bonferroni corrections reduces these thresholds to 0.00167 and 0.00333. Since our p-values for our estimates of $\tau$ and $\rho$ are well above these thresholds, we have no reason to believe the existence of any meaningful dependence between the distributions. Therefore, our assumption about goals scored and goals allowed being independent is not unreasonable.

Intuitively, the effects we described at the beginning of the section probably contribute to the slight dependence in goals scored and goals allowed. These effects, however, essentially wash out, similar to the findings in Ciccolella (2006) and Miller (2007) for baseball.

<div align="center">V. Model testing: Goodness of Fit</div>

We performed chi-squared goodness of fit tests to determine how well the Weibull densities conform to the true distributions of goals scored and goals allowed. For most teams, we tested the joint distributions by splitting our data based on the following bins:

$$[-0.5, 0.5] \cup [0.5, 1.5] \cup [1.5, 2.5] \cup [2.5, 3.5] \cup \ldots \cup [8.5, 9.5] \cup [9.5, \infty]$$

These bins are appropriate to ensure that our data occurs in the center of our bins (this is always true, as the goals scored and allowed must be non-negative integers). The number of bins was determined on a team by team basis according to each team's distribution of goals scored and goals allowed.

To perform our test, we computed the following statistics (Shao, 1999):

$$\chi^2_{GS} = \sum_{k=1}^{\#bins} \frac{\left(GS_{obs}(k) - \# \, games \int_{a_k}^{a_{k+1}} f(x; \alpha_{GS}, \gamma) \, dx\right)^2}{\# \, games \int_{a_k}^{a_{k+1}} f(x; \alpha_{GS}, \gamma) \, dx}$$

$$\chi^2_{GA} = \sum_{k=1}^{\#bins} \frac{\left(GA_{obs}(k) - \# \, games \int_{a_k}^{a_{k+1}} f(x; \alpha_{GA}, \gamma) \, dx\right)^2}{\# \, games \int_{a_k}^{a_{k+1}} f(x; \alpha_{GA}, \gamma) \, dx}$$

where $GS_{obs}(k)$ and $GA_{obs}(k)$ is the number of entries into a particular bin $k$ with left endpoint $a_k$ and right endpoint $a_{k+1}$ and

$$\# \, games \int_{a_k}^{a_{k+1}} f(x;\alpha_{GS},\gamma)dx \, / \, \# \, games \int_{a_k}^{a_{k+1}} f(y;\alpha_{GA},\gamma)dy$$ (with there being 82 games in a hockey season) is the expected proportion of the number of games a team should have in bin $k$ according to the Weibull density.

Under the null hypothesis that the distributions of goals scored and goals allowed for each particular team follow Weibull distributions, the chi-square statistics should follow a chi-squared distribution with degrees of freedom equal to one less than the total number of bines. We can reject this null hypothesis at significance level $\alpha$ if the chi-square test statistic is greater than or equal to the $(1-\alpha)^{th}$ quantile of a chi-squared distribution with degrees of freedom one less than the number of bins (Shao 2009).

Our test results are below:

Results of Chi Squared Goodness of Fit Tests – 2008/09 Season

| Team | $\chi^2_{GS}$ | Degrees of freedom | p-value | $\chi^2_{GA}$ | Degrees of freedom | p-value |
|---|---|---|---|---|---|---|
| Anaheim Ducks | 3.46 | 8 | 0.902 | 5.942 | 9 | 0.746 |
| Atlanta Thrashers | 4.084 | 9 | 0.906 | 4.699 | 9 | 0.86 |
| Boston Bruins | 4.164 | 9 | 0.900 | 2.750 | 8 | 0.949 |
| Buffalo Sabres | 4.164 | 9 | 0.9 | 2.75 | 8 | 0.949 |
| Calgary Flames | 4.447 | 8 | 0.815 | 1.058 | 8 | 0.998 |
| Carolina Hurricanes | 12.334 | 9 | 0.195 | 4.505 | 7 | 0.72 |
| Chicago Blackhawks | 7.815 | 9 | 0.553 | 6.726 | 8 | 0.566 |
| Colorado Avalanche | 9.581 | 7 | 0.214 | 10.543 | 9 | 0.308 |
| Columbus Blue Jackets | 1.713 | 8 | 0.989 | 11.238 | 8 | 0.189 |
| Dallas Stars | 7.163 | 10 | 0.71 | 9.771 | 7 | 0.202 |
| Detroit Red Wings | 13.527 | 8 | 0.095 | 13.162 | 9 | 0.155 |
| Edmonton Oilers | 12.049 | 9 | 0.211 | 9.402 | 10 | 0.494 |
| Florida Panthers | 5.783 | 9 | 0.761 | 14.589 | 8 | 0.068 |

| Team | $\chi^2_{GS}$ | Degrees of freedom | p-value | $\chi^2_{GA}$ | Degrees of freedom | p-value |
|---|---|---|---|---|---|---|
| LA Kings | 11.01 | 7 | 0.138 | 6.78 | 8 | 0.561 |
| Minnesota Wild | 10.593 | 8 | 0.226 | 8.363 | 7 | 0.302 |
| Montreal Canadiens | 9.729 | 7 | 0.204 | 4.195 | 8 | 0.839 |
| Nashville Predators | 8.104 | 8 | 0.423 | 7.517 | 9 | 0.583 |
| New York Islanders | 9.283 | 7 | 0.233 | 8.823 | 9 | 0.454 |
| New York Rangers | 9.749 | 7 | 0.203 | 8.643 | 9 | 0.471 |
| NJ Devils | 7.764 | 9 | 0.558 | 3.583 | 8 | 0.893 |
| Ottawa Senators | 7.117 | 7 | 0.417 | 4.565 | 8 | 0.803 |
| Philadelphia Flyers | 8.053 | 9 | 0.529 | 7.174 | 7 | 0.411 |
| Phoenix Coyotes | 6.872 | 7 | 0.442 | 5.177 | 8 | 0.739 |
| Pittsburgh Penguins | 7.274 | 9 | 0.609 | 8.803 | 8 | 0.359 |
| San Jose Sharks | 14.03 | 8 | 0.081 | 12.109 | 7 | 0.097 |
| St Louis Blues | 8.31 | 7 | 0.306 | 8.515 | 7 | 0.289 |
| Tampa Bay Lightning | 8.584 | 8 | 0.379 | 9.194 | 9 | 0.42 |
| Toronto Maple Leafs | 6.626 | 9 | 0.676 | 35.718 | 8 | <0.001 |
| Vancouver Canucks | 8.791 | 8 | 0.36 | 9.071 | 7 | 0.248 |
| Washington Capitals | 11.132 | 7 | 0.133 | 11.513 | 7 | 0.118 |

<u>Results of Chi Squared Goodness of Fit Tests – 2009/10 Season</u>

| Team | $\chi^2_{GS}$ | Degrees of freedom | p-value | $\chi^2_{GA}$ | Degrees of freedom | p-value |
|---|---|---|---|---|---|---|
| Anaheim Ducks | 13.052 | 8 | 0.110 | 1.105 | 8 | 0.997 |
| Atlanta | 3.862 | 8 | 0.869 | 6.736 | 8 | 0.565 |

| | | | | | |
|---|---|---|---|---|---|
| Thrashers | | | | | |
| Boston Bruins | 6.761 | 7 | 0.454 | 5.898 | 8 | 0.659 |
| Buffalo Sabres | 7.682 | 8 | 0.465 | 2.600 | 7 | 0.919 |
| Calgary Flames | 5.692 | 7 | 0.576 | 11.507 | 9 | 0.243 |
| Carolina Hurricanes | 8.613 | 9 | 0.474 | 7.056 | 8 | 0.531 |
| Chicago Blackhawks | 5.094 | 8 | 0.747 | 11.045 | 8 | 0.199 |
| Colorado Avalanche | 10.595 | 7 | 0.157 | 10.543 | 9 | 0.308 |
| Columbus Blue Jackets | 9.232 | 8 | 0.323 | 7.326 | 9 | 0.603 |
| Dallas Stars | 4.638 | 8 | 0.795 | 4.339 | 7 | 0.740 |
| Detroit Red Wings | 10.408 | 9 | 0.318 | 3.593 | 7 | 0.825 |
| Edmonton Oilers | 4.005 | 7 | 0.779 | 3.362 | 8 | 0.910 |
| Florida Panthers | 6.508 | 8 | 0.590 | 9.087 | 8 | 0.335 |
| LA Kings | 9.534 | 8 | 0.299 | 4.845 | 8 | 0.774 |
| Minnesota Wild | 1.686 | 7 | 0.975 | 2.612 | 7 | 0.918 |
| Montreal Canadiens | 8.030 | 7 | 0.330 | 5.899 | 8 | 0.659 |
| Nashville Predators | 9.005 | 8 | 0.342 | 5.672 | 8 | 0.684 |
| New York Islanders | 4.071 | 7 | 0.772 | 2.428 | 8 | 0.965 |
| New York Rangers | 4.442 | 9 | 0.880 | 6.241 | 9 | 0.716 |
| NJ Devils | 3.376 | 8 | 0.909 | 3.857 | 6 | 0.696 |
| Ottawa Senators | 3.981 | 8 | 0.859 | 4.485 | 8 | 0.811 |
| Philadelphia Flyers | 4.529 | 8 | 0.807 | 2.526 | 9 | 0.980 |
| Phoenix Coyotes | 5.160 | 7 | 0.640 | 9.152 | 7 | 0.242 |
| Pittsburgh Penguins | 9.159 | 9 | 0.423 | 4.970 | 8 | 0.761 |

| Team | $\chi^2_{GS}$ | Degrees of freedom | p-value | $\chi^2_{GA}$ | Degrees of freedom | p-value |
|---|---|---|---|---|---|---|
| San Jose Sharks | 8.608 | 10 | 0.570 | 10.245 | 9 | 0.331 |
| St Louis Blues | 3.634 | 8 | 0.889 | 7.249 | 8 | 0.510 |
| Tampa Bay Lightning | 3.190 | 8 | 0.922 | 5.179 | 9 | 0.818 |
| Toronto Maple Leafs | 8.380 | 7 | 0.300 | 8.373 | 8 | 0.398 |
| Vancouver Canucks | 9.182 | 9 | 0.421 | 5.833 | 9 | 0.757 |
| Washington Capitals | 8.488 | 8 | 0.387 | 5.847 | 7 | 0.558 |

Results of Chi Squared Goodness of Fit Tests – 2010/11 Season

| Team | $\chi^2_{GS}$ | Degrees of freedom | p-value | $\chi^2_{GA}$ | Degrees of freedom | p-value |
|---|---|---|---|---|---|---|
| Anaheim Ducks | 2.129 | 8 | 0.977 | 8.815 | 9 | 0.455 |
| Atlanta Thrashers | 3.798 | 8 | 0.875 | 9.083 | 10 | 0.524 |
| Boston Bruins | 17.084 | 9 | 0.047 | 4.434 | 8 | 0.816 |
| Buffalo Sabres | 3.855 | 9 | 0.921 | 4.679 | 8 | 0.791 |
| Calgary Flames | 3.844 | 9 | 0.921 | 8.747 | 8 | 0.364 |
| Carolina Hurricanes | 10.240 | 8 | 0.249 | 16.257 | 9 | 0.062 |
| Chicago Blackhawks | 3.419 | 8 | 0.905 | 6.856 | 7 | 0.444 |
| Colorado Avalanche | 6.993 | 8 | 0.537 | 15.457 | 8 | 0.051 |
| Columbus Blue Jackets | 7.354 | 7 | 0.393 | 8.382 | 8 | 0.397 |
| Dallas Stars | 7.542 | 7 | 0.375 | 6.796 | 8 | 0.559 |
| Detroit Red Wings | 4.918 | 8 | 0.766 | 6.881 | 8 | 0.550 |
| Edmonton Oilers | 3.536 | 8 | 0.896 | 9.956 | 9 | 0.354 |
| Florida Panthers | 6.982 | 8 | 0.539 | 15.389 | 6 | 0.017 |
| LA Kings | 10.224 | 7 | 0.176 | 8.336 | 8 | 0.401 |
| Minnesota Wild | 4.702 | 7 | 0.696 | 6.327 | 9 | 0.707 |
| Montreal Canadiens | 19.026 | 9 | 0.025 | 5.528 | 9 | 0.786 |
| Nashville Predators | 8.597 | 7 | 0.283 | 5.222 | 7 | 0.633 |
| New York Islanders | 3.660 | 9 | 0.932 | 5.538 | 8 | 0.699 |
| New York Rangers | 5.027 | 9 | 0.832 | 10.226 | 7 | 0.176 |
| NJ Devils | 4.906 | 7 | 0.671 | 6.936 | 8 | 0.544 |
| Ottawa Senators | 6.791 | 7 | 0.451 | 8.610 | 8 | 0.376 |
| Philadelphia Flyers | 4.603 | 9 | 0.867 | 57.942 | 8 | 0.000 |
| Phoenix Coyotes | 7.667 | 7 | 0.363 | 13.447 | 8 | 0.097 |
| Pittsburgh Penguins | 4.262 | 9 | 0.893 | 6.205 | 8 | 0.624 |
| San Jose Sharks | 10.259 | 7 | 0.174 | 7.808 | 7 | 0.350 |
| St Louis Blues | 5.978 | 9 | 0.742 | 8.638 | 9 | 0.471 |
| Tampa Bay Lightning | 6.552 | 9 | 0.684 | 6.020 | 9 | 0.738 |
| Toronto Maple Leafs | 12.819 | 8 | 0.118 | 6.665 | 8 | 0.573 |
| Vancouver Canucks | 7.742 | 8 | 0.459 | 9.182 | 8 | 0.327 |
| Washington Capitals | 10.289 | 7 | 0.173 | 6.928 | 7 | 0.436 |

Our p values are almost always well above commonly accepted critical thresholds of 0.05 and 0.10. Furthermore, after instituting Bonferroni corrections, our critical thresholds drop to 0.00167 and 0.00333 respectively, and all of our distributions except the Toronto Maple Leafs GA in 2008/09 and the Philadelphia Flyers GA in 2010/11 fall below our necessary critical thresholds. As a result, it is not unreasonable to assume that virtually all of our distributions of GS and GA for each of our 30 teams follow Weibull distributions.

## VI. Conclusions and Future Research

Our results provide statistical justification for applying the Pythagorean Won-Loss formula to hockey. We estimate $\gamma$ via maximum likelihood estimation to be slightly above two. Our tests of statistical independence and goodness of fit are quite strong, illustrating that the Pythagorean Won-Loss formula is just as applicable to hockey as it is to baseball. We hope this research encourages the use of the Pythagorean Won-Loss formula as an evaluative tool in hockey. There are a number of potential avenues of future research that we hope this work will encourage:

1. Future research should go on to examine the statistical appropriateness of applying the Pythagorean Won-Loss formula to other sports, such as basketball and soccer. Researchers could then use the formula as a basis for comparing teams of different eras and understanding the effects of hiring well-known coaches or superstars, as well as the expected gains resulting from mid-season signings.

2. One could also perform a more micro analysis as suggested in Miller (2006) to incorporate lower order effects. Baseball has several natural candidates, ranging from park effects to the presence or absence of a designated hitter depending on where the game is played. Similarly, there are natural candidates to investigate in hockey. The first is rink effects, ranging from having the home crowd to slight differences in the rinks (see Weiner 2009 for some of the differences between rinks, even though they all have the same dimensions for the ice). Other items include power plays (which means both how well a team does on power plays, as well as how likely they or the opponent is to provide an opportunity), "meaningless" goals late in the game (such as goals scored by the leading team when the trailing team pulls its goalie), and overtime scoring (and its relation to classifying the game as a win or a loss). As our model already does a great job explaining the data, it is likely that these are lower order effects that mostly wash out, but it would still be interesting to see the size of their effects.

3. Almost surely professional sports players do not discuss how to ensure their scoring conforms to a Weibull distribution. Regardless, we used such a model here as doing so leads to a tractable double integral that can be solved in closed form. One of primary advantages of the Pythagorean formula is the simplicity of the resulting statistic; however, in an age of powerful and ever-present computing power, the need for a simple statistic is lessened. Consequently, there are several other approaches one may take:

   a. One possibility is to look at linear combinations of Weibull distributions. The resulting fit to the data cannot be worse, as our situation is just the special case of one Weibull distribution. One would have a sum of individually tractable integrals, all yielding closed-form expressions.
   b. Along these lines, one could replace a Weibull distribution with a linear combination of a Weibull distribution and a point mass at zero. Such a model allows one to accommodate for the probability of being shut out and have another density to model scoring. A similar idea is used via a quasi-geometric model in (Glass and Lowry, 2008) to model scoring in baseball games.

c. The scoring data for both baseball and hockey is well-modeled by a one-hump distribution, namely the probability initially rises to a maximum and then continuously falls. Instead of using a Weibull distribution, one could use a Beta distribution instead, where the

density becomes $f(x; a, b) = \dfrac{\Gamma(a+b)}{\Gamma(a)\Gamma(b)} x^{a-1}(1-x)^{b-1} I(0 \le x \le 1)$

(with a, b > 0 our shape parameters); here $\Gamma$ is the Gamma function (which is a generalization of the factorial function, with $\Gamma(n+1) = n!$ for n a non-negative integer) and $I(0 \le x \le 1)$ is the indicator function which is 1 for x between 0 and 1 and 0 otherwise. For many choices of a and b we find that a Beta distribution captures the general shape of the observed scoring data; however, while closed-form expressions exist for the mean and the variance of the Beta distribution in terms of its parameters, for general choice of the parameters we do not have a nice closed form expression for the needed double integral. Thus, if Beta distributions were to be used, one would be reduced to numerical approximations to find the dependence of the winning percentage on the parameters of the teams.

## Acknowledgements

We would like to thank Eric Fritz for writing a Java script to download hockey data from ESPN.com, Mike Taylor, Hovannes Abramyan and Arnal Dayaratna for proofreading, Armaan Ambati for help with the data and Benjamin Kedem for insightful comments.

## VII.  References

American Institute of Physics, Streamlining The 'Pythagorean Theorem Of Baseball' (2004, March 30), ScienceDaily. http://www.sciencedaily.com/releases/2004/03/040330090259.htm.

Apple, Chris and Foster, Marc.  "Playoffs projections based on Pythagorean Performance"  2002, January 10.  *Sports Illustrated*. http://sportsillustrated.cnn.com/statitudes/news/2002/01/09/just_stats/

Casella G. and Berger R., Statistical Inference, Second Edition, Duxbury Advanced Series, 2002.

Ciccolella, R. *Are Runs Scored and Runs Allowed Independent*, By the Numbers **16** (2006), no. 1, 11–15.

Cochran, J. and Blackstock, R. *Pythagoras and the National Hockey League*, Journal of Quantitative Analysis in Sports (2009), **5** (2), Article 11.

ESPN. "NHL Expanded Standings — 2008/09." ESPN: The Worldwide Leader in Sports. http://espn.go.com/nhl/standings/_/type/expanded/year/2009.

ESPN. "NHL Expanded Standings — 2009/10." ESPN: The Worldwide Leader in Sports. http://espn.go.com/nhl/standings/_/type/expanded/year/2010.

ESPN. "NHL Expanded Standings — 2010/11." ESPN: The Worldwide Leader in Sports. http://espn.go.com/nhl/standings/_/type/expanded/year/2011.

Foster, Mark. "Benchmarcs, Performing to Expectations" (2010, December 19). *Hockey Prospectus* http://www.hockeyprospectus.com/article.php?articleid=715

Glass, D. and Lowry, P. J. *Quasigeometric Distributions and Extra Inning Baseball Games*, Mathematics Magazine (2008), **81** (2), 127-137.

Hogg, R.V.; Craig, A.T.; and McKean, J.W., Introduction to Mathematical Statistics, Sixth Edition, Prentice Hall Inc, 2004.

James, B. The Bill James Abstract, self-published, 1979.

James, B. The Bill James Abstract, self-published, 1980.

James, B. The Bill James Abstract, self-published, 1981.

James, B. The Bill James Abstract, Ballantine Books, 1982.

James, B. The Bill James Abstract, Ballantine Books, 1983.

Miller, S.J., *A Derivation of the Pythagorean Won-Loss Formula in Baseball*, Chance Magazine **20** (2007), no. 1, 40–48. An abridged version appeared in The Newsletter of the SABR Statistical Analysis Committee **16** (February 2006), no. 1, 17–22, and an expanded version is available at http://arxiv.org/abs/math/0509698.

Oliver, D. Basketball On Paper, Potomac Books, 2004.

Rosenfeld, J.W.; Fisher, J.I.; Adler, D; and Morris, C. *Predicting Overtime with the Pythagorean Formula*, Journal of Quantitative Analysis in Sports (2010), **6**, no. 2.

Ryder, A. *Win Probabilities: a tour through win probability models for hockey* (2004), Hockey Analytics http://www.hockeyanalytics.com/Research_files/Win_Probabilities.pdf.

Schatz, A. *Pythagoras on the Gridiron*. Football Outsiders, July 14, 2003. http://www.footballoutsiders.com/stat-analysis/2003/pythagoras-gridiron.

Shao, J. Mathematical Statistics, Springer, 1999.

Weiner, E. *Not every 200 foot by 85 foot NHL rink is the same*, Off the Wall, October 9, 2009 (5:00pm). http://www.nhl.com/ice/news.htm?id=501626.